\documentclass[aps,prl]{revtex4}  
\usepackage{graphicx}
\usepackage{epsfig}
\begin{document}
\title{Crucial stages of protein folding through a solvable
model: predicting target sites for enzyme-inhibiting drugs.}

\author{Cristian Micheletti, Fabio Cecconi,  Alessandro Flammini and Amos Maritan\\
\small International School for Advanced Studies (SISSA/ISAS),
Trieste, ITALY}
\date{\today} 

\newpage

\begin{abstract}
An exactly solvable model based on the topology of a protein native
state is applied to identify bottlenecks and key-sites for the folding
of HIV-1 Protease. The predicted sites are found to correlate well
with clinical data on resistance to FDA-approved drugs. It has
been observed that the effects of drug therapy are to induce multiple
mutations on the protease.  The sites where such mutations occur
correlate well with those involved in folding bottlenecks identified
through the deterministic procedure proposed in this study. The high
statistical significance of the observed correlations suggests that
the approach may be promisingly used in conjunction with traditional
techniques to identify candidate locations for drug attacks.
\end{abstract}

\maketitle
\section{Introduction}

One of the open fundamental questions in molecular biology is how to
predict the folded state of a protein from the knowledge of its
sequence. Despite a large increase in available computing power in the
past years, it has been impossible to answer this question by means of
computer simulations of various degrees of complexity and detail.
However an increasing amount of
experimental~\cite{Fersht95,sh3A,sh3B,Chiti99,Plaxco} and
theoretical results~\cite{Miche99b,M00,baker,Clem,cieplak} supports
the view that the folding of natural proteins into their native state
is largely influenced by the native-state topology (for a brief review
see~\cite{bakernature}). Accordingly, the folding process
is regarded as a well defined sequence of obligatory steps to be taken
in order to reach the native state. Even if protein sequences have
evolved to fold efficiently, the kinetics en-route to the native state
might be hindered by the realization of particularly difficult
(rate-limiting) steps, such as the formation of non-local amino acid
interactions (contacts) that usually requires the overcoming of
large entropy barriers. Some non-local native contacts
are rather crucial for the folding process, because their
formation helps establishing further native interactions and leads to
a rapid progress along the folding pathway until another barrier is
met. Their formation is associated to bottlenecks for the entire
folding process. Strikingly, the amino acids involved in such
crucial contacts are those for which the largest changes in the folding
kinetics are observed in site-directed mutagenesis
experiments~\cite{Fersht95}, as first proven for CI2 and
Barnase~\cite{Miche99b}. This suggests that protein sequences have
been carefully optimised so to exploit the conformational entropy
reduction accompanying the folding process \cite{Funnel2} through the
selection of the key amino acids. The number and importance of
bottlenecks depends significantly on several factors. Among the most
important are the contact order of the protein\cite{baker} and whether
it folds in two or more stages~\cite{Jackson}.

In previous studies~\cite{hiv,Settanni}, we have shown how the most
delicate folding stages can be identified within a molecular dynamics
approach, by monitoring the formation probability of native and
non-native contacts from the unfolded to the native state. This can
either be done as a function of time at a fixed temperature around the
folding temperature or working at thermal equilibrium for a succession
of decreasing temperatures (annealing). In principle, the two
approaches need not to be equivalent but, for the quantities we have
investigated, they give consistent results.  Then, concerning the
indentification of crucial contacts, one can safely concentrate on
studying thermodynamic equilibrium at various temperatures. The main
limitation of Molecular Dynamics (MD) and MonteCarlo (MC) simulations,
especially for long protein chains, is that they are extremely
time-demanding and plagued with statistical errors which can affect
the predictions based on the study of the relative sensitivity of
contact formation.  Therefore it would be highly desirable to develop
a suitable theoretical model, amenable to a deterministic (and
computationally fast) treatment, thus resulting in a deeper
understanding of the problem. Ideally, such a model should encompass
all the ``necessary ingredients'' that usually are included in
computer simulations: peptide chain constraints, effective
interactions between residues, favourable monomeric positions, etc. In
the following we describe a recently developed theoretical scheme
\cite{gaussian}, that, while being very simplified and approximate
compared to other schemes based on MD or MC simulations, can be treated
analytically, leading to expressions that can be evaluated
exactly. The calculated quantities rival those obtained through more
sophisticated but computationally demanding MC and MD techniques.  The
purpose of the present paper is to show how the model can be employed
to yield helpful observables to identify the folding bottlenecks.  In
particular we apply the method to the HIV-1 Protease (HIV-1 PR), an
enzyme which is crucially involved in the HIV infection \cite{condra}.
In general, the accurate knowledge of bottlenecks has important
pharmaceutical ramifications because their knowledge may be exploited
in a rational drug design.  Due to the large amount of available
clinical data, HIV-1 PR is a natural choice for a stringent test of
our automated predictive scheme.

\section{Theory}

 The model we adopt builds on the importance of the native state
topology in steering the folding process, that is in bringing into
contact pairs of amino acids that are found in interaction in the
native state. A primary quantity of interest that we shall calculate
is the probability that a given native contact is established at a
definite stage of the folding process. Probably, the oldest attempt to
calculate such quantity dates back to Flory who tried to estimate the
probability $p_{ij}$ that two sites $i$ and $j$ in a long harmonic
chain (the peptide) are in contact \cite{flory}. The approximation
introduced by Flory was to neglect correlations between residues,
which amounts to considering the chain embedded in a
highly-dimensional space. As a result, the $p_{ij}$'s are a decreasing
function of the sequence separation $|i-j|$. Clearly, this
approximation is not apt to pinpoint the key folding sites, since it
exploits the native topology at the simplest level; in fact it takes
into account only the contact order of native interactions. The Flory
approach, however, can be refined by incorporating correlations
between the formation of pairs, triplets etc. of contacts
\cite{chan90,goddard,CT2}. Here we use a recently introduced energy
function that allows to calculate the $p_{ij}$'s within a
self-consistent analytic scheme. The strategy is similar in spirit to
that of Go and Scheraga \cite{Go} where only the formation of native
interactions is energetically rewarded and is common to all recent
approaches which exploits the native state topology
~\cite{Miche99b,M00,baker,Clem,cieplak}.

We describe the proteins by the coordinates ${\bf r}_i$ of the
C$_{\alpha}$ atom of the $i$-th amino acids. The simplified energy
functional for the chain of $N$ residues is
\begin{equation}
H = \frac{K T}{2} \sum_{i=1}^{N-1}
({\bf r}_{i,i+1} - {\bf r}^0_{i,i+1})^2 + \frac{1}{2} \sum_{i \neq j}
\Delta_{ij} [({\bf r}_{ij} - {\bf r}^0_{ij})^2 - R^2] \theta_{ij}
\label{eq:gauss}
\end{equation}
where $K$ is the strenght of the peptide bonds, assumed to be
harmonic, and $T$ is the absolute temperature in units of the Boltzmann
constant.

The relative position between amino acid centroids is denoted by ${\bf
r}_{ij} = {\bf r}_i - {\bf r}_j$ and the corresponding native
positions are indicated with the superscript $0$. $\Delta$ is the
contact matrix, whose element $\Delta_{ij}$ is 1 if residues $i$ and
$j$ are in contact in the native state (i.e. their C$_\alpha$
separation is below the cutoff $c=6.5$ \AA) and 0 otherwise. The
matrix $\Delta_{ij}$ along with the set ${\bf r}^0_{ij}$ encodes the
topology of the protein.  The factor $\theta_{ij}$ has the form
\begin{equation}
\theta_{ij} = \Theta(R^2 - ({\bf r}_{ij} - {\bf r}^0_{ij})^2)
\label{eq:tetaij}
\end{equation}
where $\Theta(x)$ is the unitary step function and $R$ is a distance
cutoff defining the range of the interaction between non consecutive
amino acids. In standard off-lattice approaches, the interaction
$V(d)$ between non-bonded amino acids at a distance $d$, is taken to
be a square well potential, or some type of Lennard-Jones interaction.
Our choice in Eq. (\ref{eq:gauss}) is a sort of ``harmonic well''
which, while being physically sound and viable, is suitable for a
self-consistent treatment, as explained below. The location of the
outer rim of the well is controlled by $R$, which can be set to a few
Angstroms ($R = 3$ \AA\ in the present study) to penalise
conformations where the the separation of two residues differs too
much from the native one. In the native state each $\theta_{ij}$ is
close to $1$ while in the denaturated state case $\theta_{ij}$ are
usually negligible.

While the present form of the model does not accurately describe the
effects of self-avoidance this does not lead to a qualitatively wrong
behaviour in the highly-denatured ensemble (large $T$). The treatment
of steric effects becomes progressively more accurate as temperature
is lowered. In fact, the model guarantees that the native state is the
true ground state and therefore protein conformations found at low
temperature inherit the native self-avoidance. The connectedness
of the chain, as well as its entropy, are captured in a simple but
non-trivial manner. The most significant advantage of the model is
that it can be used to explore the equilibrium thermodynamics without
being hampered by inaccurate or sluggish dynamics.

Two limit cases of model \ref{eq:gauss} are worthy of notice. In the
absence of any bias towards the target structure (i.e. when both
$\Delta_{ij}$ and the $\{r^0\}$'s are removed) the model reduces to the
standard gaussian polymer model whose behaviour is exactly known
\cite{flory,Klo99}.  Furthermore, the limit when $T \to 0$ (when all
native contacts are established and the bonded-energy term
fluctuations are negligible) the model reduces to the gaussian network
model that has been introduced and used to study
the near-native vibrational properties of several proteins
\cite{bah97,hiv99,kes00,ani01}.

The thermodynamics of the model is fully determined by the partition
function
\begin{equation}
{\cal Z}(T) = \int \prod_{i=1}^N d^3 r_i \exp(-H/T).
\label{eq:Z}
\end{equation}

In the integral of equation (\ref{eq:Z}) and in the following, it is
always meant that translational invariance is always explicitly broken
by fixing, for example, the centre of mass of the system (see Appendix).

The integral~(\ref{eq:Z}) is still hard to treat
analytically, due to the presence of non-quadratic interactions in the
last term of Hamiltonian~(\ref{eq:gauss}). We thus perform a further,
but non-trivial, simplification by replacing $H$ with the variational
hamiltonian $H_0$
\begin{equation}
H_0 = \frac{K T}{2} \sum_{i=1}^{N-1}
({\bf r}_{i,i+1} - {\bf r}^0_{i,i+1})^2 + \frac{1}{2} \sum_{i \neq j}
\Delta_{ij} [({\bf r}_{ij} - {\bf r}^0_{ij})^2 - R^2] p_{ij}
\label{eq:H_0}
\end{equation}
where the factors $\theta_{ij}$ are now subtituted by parameters
$p_{ij}$ independent of the coordinates. Due to its quadratic form,
the model described by Eq. (\ref{eq:H_0}) can be solved with the
standard techniques for Gaussian integrals. Such parameters have to
be optimally determined so to ensure self-consistency:
\begin{equation} p_{ij} = \langle \Theta(({\bf
r}_{ij} - {\bf r}^0_{ij})^2 - R^2) \rangle_0.
\label{eq:self}
\end{equation}
The symbol $\langle...\rangle_0$ indicates that the
thermal averages are performed through the Hamiltonian $H_0$. Now in
such self consistent approach the problem is fully solved and we can
compute the resulting partition function from which we extract all the
thermal properties and averages. In particular the logarithm of the
partition funzion, $\ln ({\cal Z})$,
has the following explicit expression:

\begin{equation}
\ln ( {\cal Z}) = {3(N-1) \over 2} \ln (2 \pi) - {3 \over 2} \ln N - {3 \over 2} \ \ln
(\det \null^\prime {M}) +  {R^2 \over 2 T} \sum_{lm} \Delta_{lm} \ p_{lm}
\label{eqn:lnz}
\end{equation}

\noindent where the matrix $M$ is defined as:

\begin{equation}
{{M}}_{i,j} = \left \{
\begin{array}{r  l}
K (2 - \delta_{i,1} - \delta_{i,N}) + 2 \sum_l \Delta_{i,l}\, p_{i,l}/T \
\ &  {\rm for }\  i=j \\
& \\
-2 p_{i,j} \Delta_{i,j}/T - K \left[ \delta_{i,j+1} +
\delta_{i,j-1}\right] \ \ & {\rm for }\ i \not= j \ .
\end{array}
\right.
\end{equation}

\noindent and the prime in (\ref{eqn:lnz}) denotes that the zero
eigenvalue of $M$ has to be omitted (see Appendix).

The quantities $p_{ij}$ in Eq.~(\ref{eq:self}) represent precisely the
occurrence probability of a contact between residues $i$ and $j$ and
indicate the frequency with which that native contact is established.
At thermal equilibrium their dependence on temperature reflect the
status of compactness of the protein molecule. For instance, well
below the folding temperature, $T_F$, each $p_{ij}(T)$ is expected to
assume a value close to unity, as all native contacts are already
formed. Instead, for temperatures much larger than $T_F$, all
$p_{ij}(T)$ tend to be very small, reflecting the low propensity of
the protein to establish contacts.

Thermodynamics quantities can be easily derived from the $p_{ij}$'s.
Another quantity necessary to characterize the folding transition is
the specific heat, which exhibits one or more peaks in correspondence
of significant structural rearrangements of the protein
conformation. Since every energy change is mainly associated to the
formation of native interactions, we address the question of which
native contacts contribute mainly to the peak(s) of the specific
heat. A clear answer to this question is readily found in the
temperature behaviour of frequencies $p_{ij}$. Indeed, each
$p_{ij}(T)$ exhibits a sigmoidal shape, and the modulus of its
derivative develops a sharp maximum in correspondence of the point of
inflection (crossover temperature). The importance of every native
contact $i-j$ turns out to be characterized by the crossover
temperature and the maximum slope of its $p_{ij}$, which can be
regarded as an indicator of its degree of cooperativity. In fact, the
most important contacts are those with high crossover temperature and
associated high cooperativity.

This fact allows a complete identification and classification of the
bottlenecks, because we are now able to indentify those contacts that
are termodinamically relevant to peaks and shoulders of the specific
heat.

\section{Application to HIV-1 Protease}

The human immunodeficiency virus (HIV) encodes a protease, HIV-1 PR,
whose inhibition is crucial to prevent the maturation of infectious
HIV particles~\cite{condra}.  The role of the Protease in the
infection spreading is to act as "molecular scissor" cleaving inactive
viral polyproteins into smaller, functional proteins.  In the presence
of protease inhibitors, viral particles are unable to mature and are
rapidly cleared. Extensive clinical trials have lead to the
development of five HIV-1 PR inhibitors approved by the
Food and Drug Administration (FDA):
Saquinavir mesylate (SAQ), Ritonavir (RIT), Indinavir sulfate
(IND), Nelfinavir mesylate (NLF) and Amprenavir (APR) \cite{BIOCH88}.
Such drugs are particularly effective in short-term treatments, while
their long-term efficacy is limited by resistance. Indeed mutants
resistant to protease-inhibitors can emerge in vivo already after less
than one year~\cite{condra}. Table \ref{tab:res} summarises the list
of HIV-1 PR known mutating sites causing drug resistance.

In an earlier work, the study of the near-native harmonic vibrations
of the HIV-1 PR has shown that a number of sites that are paramount to
the stability of the native enzyme are close to some of the residue of
Table 1 \cite{hiv99}. The self-consistent scheme of eqn. \ref{eq:H_0}
allows to extend this result by modelling the partially-folded
ensemble at finite temperature.

In particular, we will be concerned in the characterization of such
ensemble near the folding transition temperature. The motivation to do
so stems from a recent study~\cite{hiv} where we have shown that such
mutating amino acids correspond, with high statistical significance,
to sites involved in the folding kinetic bottlenecks. The rationale
for this finding is that the most effective drugs can be eluded only
by mutations occurring in correspondence of the key sites. Due to the
sensitivity of the folded native conformation to these sites, only
fine-tuned mutations are allowed in correspondence of these sites.
Such mutation have to result in a native-like enzymatic activity and
in the avoidance of the drug action. These constraint act as a severe
selective pressure on the mutated proteases that the HIV virus is able
to express. As a result, the mutations that will ultimately cause
drug-resistence are expected to occur in correspondence of the crucial
sites. These residues are heavily influenced by the native topology,
and hence should display little dependence on the particular
(effective) drug to be eluded.

It is therefore our purpose to apply the scheme introduced in the
previous section and identify the key residues within our
topology-based scheme. The method, being completely analytic, is free
from statistical uncertainty, common to all MC and MD simulation
methods, or from difficulty (due to spatial restraints) to reach the
target native state below the folding temperature.

\section{Results and discussion}

\begin{figure}
\begin{center}
\psfig{figure=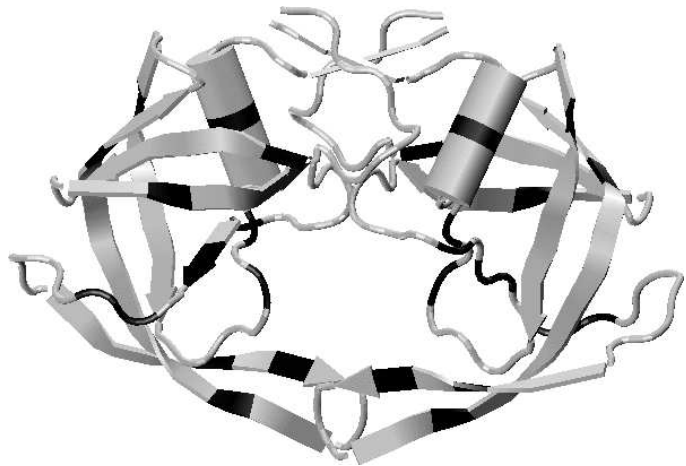,width=0.5\textwidth}
\end{center}
\caption{Structure of HIV-1 PR dimer \protect\cite{condra}. The
highlighted locations indicate residues where mutations causing
drug-resistance are observed.}
\label{fig:dimero}
\end{figure}

\begin{figure}
\begin{center}
\psfig{figure=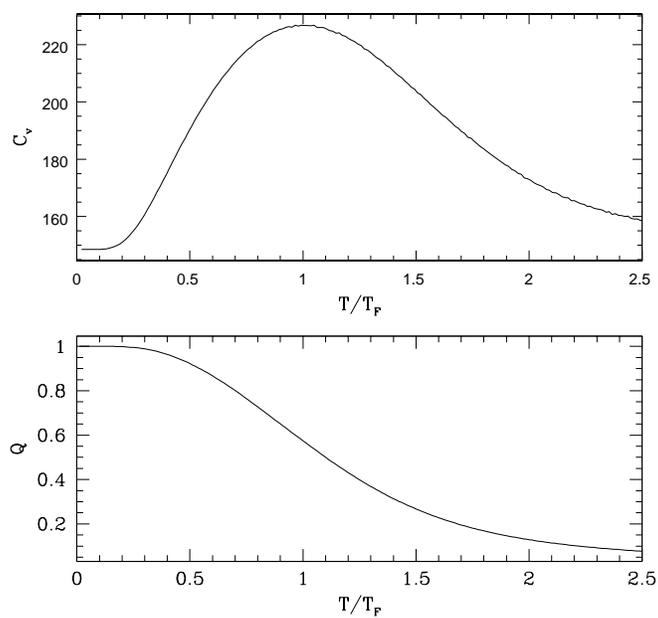,width=0.5\textwidth}
\end{center}
\caption{Specific heat and overlap of a monomer of the HIV-1 PR. The
temperature is scaled with the temperature $T_F$ where the specific heat
peak occurs} \label{fig:C} \end{figure}

The structural model at the basis of our analysis is the free enzyme
\cite{condra}. It is a homodimer with C2 symmetry, each subunit being
composed by 99 residues (Fig.~\ref{fig:dimero}).  Previous
studies~\cite{hiv} have shown that geometrically important residue
positions can be obtained considering a single monomer. Indeed the
specific heat of the whole homodimer on decreasing the temperature
shows a peak in correspondence of the folding of each sub-unit and
then at lower temperature another peak signals the aggregation of the
two sub-units. Thus, in the following, we will be concerned only with
a single monomer.  The specific heat is obtained through
numeric differentiation of the average internal energy, which has the
following explicit analytic expression in terms of the $p_{ij}(T)$'s
and the quantities introduced before:

\begin{equation}
\langle E \rangle = { 3\, (N-1)\, T\over 2} - {R^2 \over 2} \sum_{ij}
\Delta_{ij}\, p_{ij}(T)\ .
\label{eq:avg_e}
\end{equation}
\noindent The study of Go and Scheraga \cite{Go} showed that systems
described by energy-scoring-functions that reward the formation of
native contacts display cooperative (all-or-none) folding transitions
with an associated peak(s) in the specific heat. Consistently with
these expectations, the specific heat calculated by differentiating
Eq. (\ref{eq:avg_e}) with respect to $T$ shows a single peak, see
Fig.~\ref{fig:C}, thus providing an unambiguous criterion for
identifying the folding transition temperature, $T_F$. The width of
the specific heat peak at the folding transition in Figure~\ref{fig:C}
is larger than the typical one found in experimental \cite{Jackson}
and theoretical studies \cite{Kaya1,kaya2}. It is possible to enhance the
cooperativity of the transition by intervening on the actual value of
$K$ in Eq. (\ref{eq:gauss}); in fact, a decrease of $K$ leads to
sharper transitions. An alternative criterion for fixing the value of
$K$ is provided by its influence on the average amount of
native structure that is formed at the native state. Since we are
particularly interested in monitoring the progressive establishment of
native contacts, we adopted this second possibility to set the value
of $K$. In fact, by choosing $K=1/15$ in (\ref{eq:gauss}), we ensure
that, at $T_F$, the average fractional occupation of native contacts,
$q$:

\begin{equation}
q = {\sum_{i,j}^\prime \Delta_{i,j} \ p_{i,j} \over \sum_{i,j}^\prime \Delta_{i,j}}
\end{equation}

\noindent is about 50 \% (see Fig.~\ref{fig:C}), as established in
several experiments and numerical studies. The primed summation symbol
indicates that the sum is not carried out over consecutive pairs. The
degree of native similarity, $q$ is a useful overall indicator to
monitor the progress towards the native state in a folding process
\cite{CT,Funnel5}.  While the ultimate quantities of interest are the
$p_{i,j}$'s, it is useful to consider an intermediate level of
description and focus on the whole network of contacts that a given
site takes part to. A natural order parameter is provided by the
``average environment formation'' \cite{Finkel,karp} which, for a
generic site $i$ is defined as:

\begin{equation}
P_i = {\sum_{j}^\prime \Delta_{i,j} \ p_{i,j} \over\sum_{j}^\prime \Delta_{i,j} }\;
.
\label{eq:phi}
\end{equation}

\begin{figure}
\begin{center}
\psfig{figure=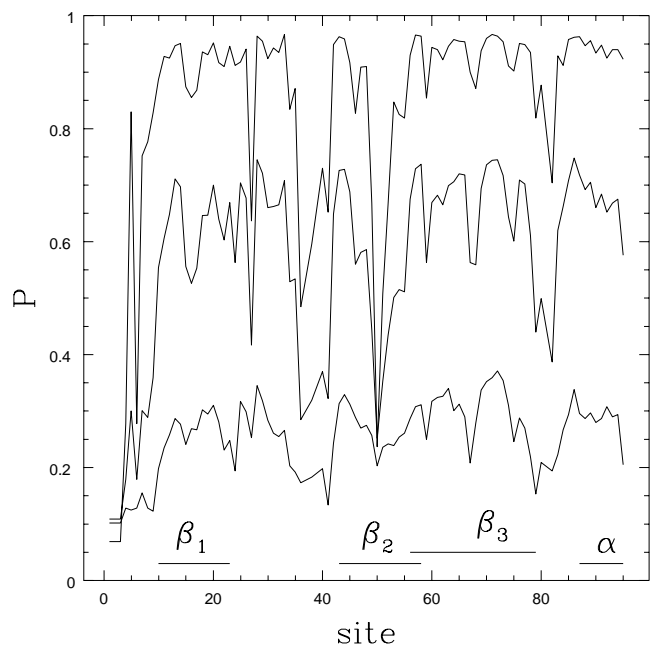,width=0.5\textwidth}
\end{center}
\caption{ Plot of $P_i$, the degree to which amino acid $i$ is in a
native-like conformation, versus $i$.  In ascending order the curves
are calculated at $T/TF$= 1.5, 1.0 and 0.5. The bar at the bottom
shows the secondary structure associated with amino acid $i$.}
\label{fig:D}
\end{figure}

\begin{figure}
\begin{center}
\psfig{figure=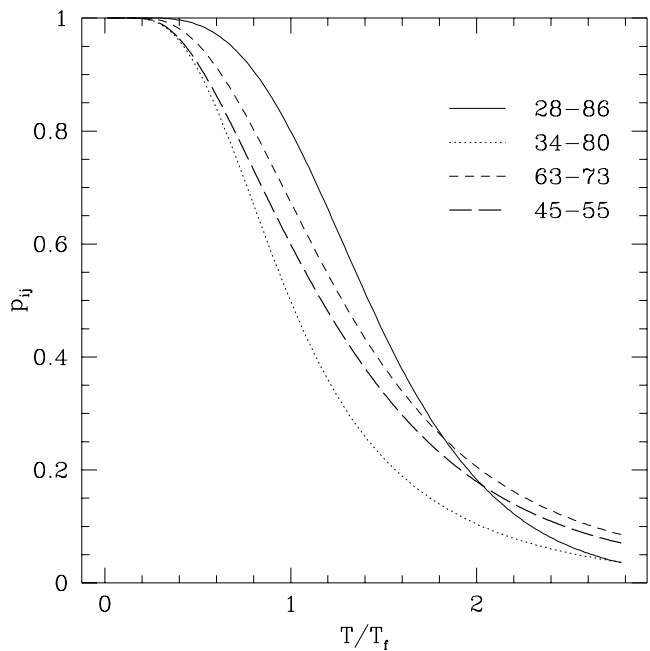,width=0.5\textwidth}
\end{center}
\caption{Typical behaviour of contact probabilities, $p_{i,j}$ versus
$T/T_F$ for four native contacts involving pair of sites with different sequence separation
and degree of native burial.}
\label{fig:curves}
\end{figure}

\noindent $P_i$ is a measure of the fraction of established native
contacts the $i$-th residue partecipate to (clearly, $P_i$ is defined
only when the denominator of eqn. \ref{eq:phi} is non-zero). The
environment profiles for three different temperatures are shown in
Fig.~\ref{fig:D}. The irregular behaviour of the profiles results
from a complex interplay of the burial of the sites and the locality
of their contacts. The hierarchical formation of secondary structures
at high temperature is clearly visible. It is instructive to correlate
the location of the sites known to cause resistance to drug treatments
(see Table~\ref{tab:res}) with the features of the profiles. In
particular, several mutating sites responsible for drug resistance
(see Table \ref{tab:res}) can be found in correspondence of the peaks
of the environments (see in particular sites 20,63,71,77,84). The most
precise way to identify the key residues is, however, through the
analysis of the fractional occupation of native contacts and not
through the environments, since they only carry averaged
information. Typical $p_{i,j}$ curves as a function of temperatures
are shown in Fig.~\ref{fig:curves}.

\begin{figure}
\begin{center}
(a)\psfig{figure=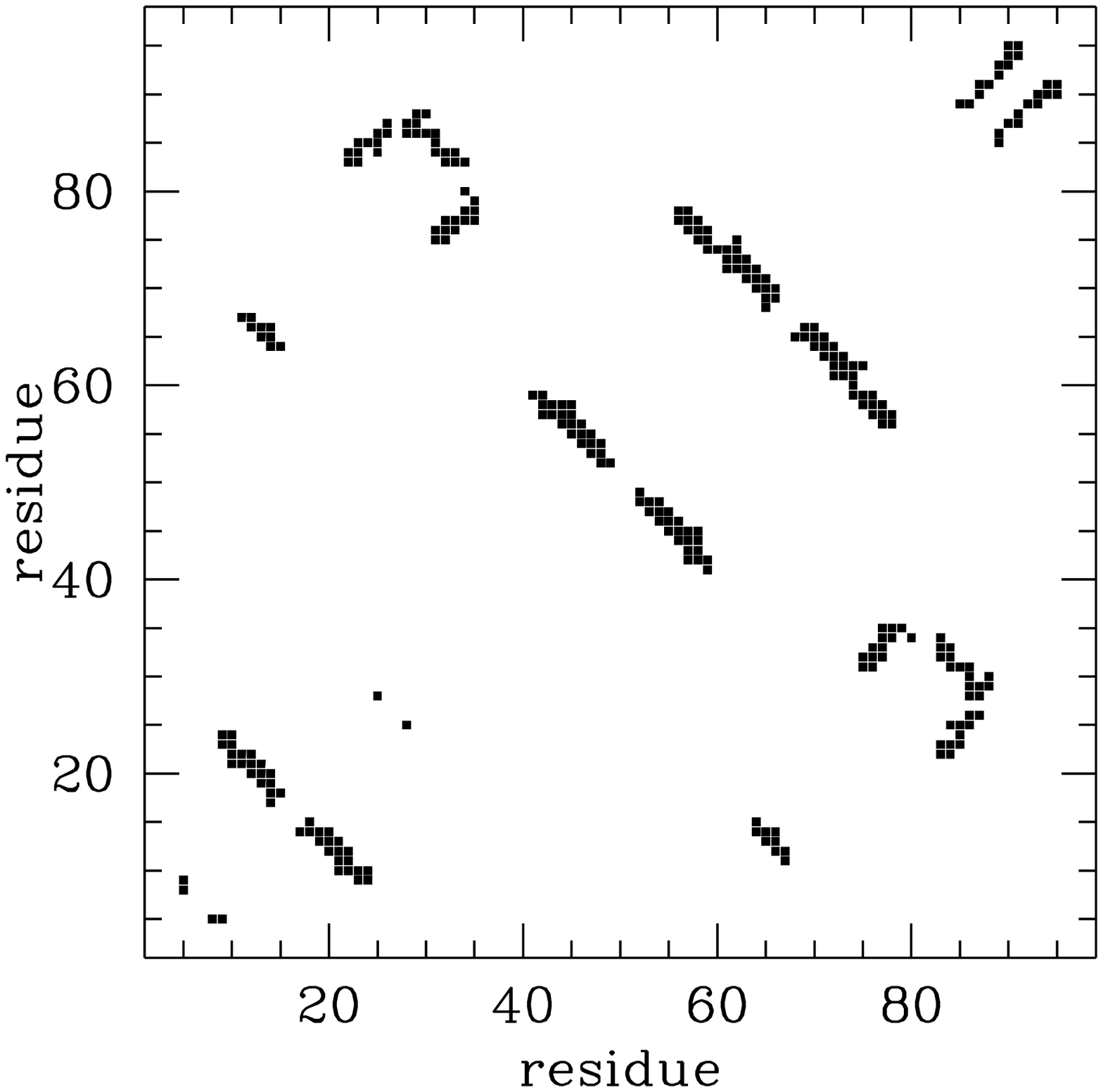,width=0.46\textwidth}
(b)\psfig{figure=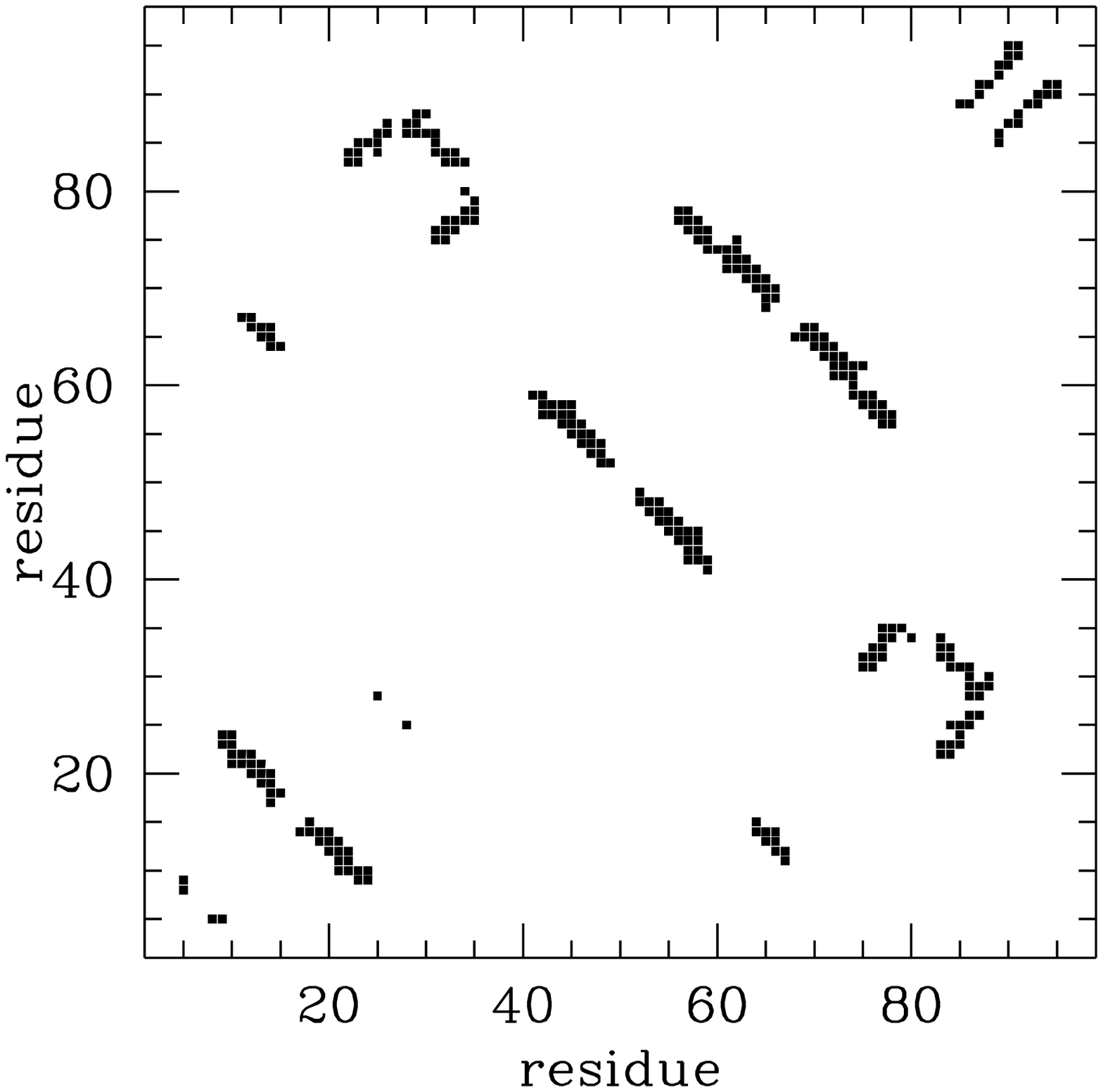,width=0.46\textwidth}
\end{center}
\caption{Color-coded contact map of HIV-1 PR monomer. (a) Contacts with a
large [small] crossover temperature are shown in red [blue].(b) Contacts with a
large [small] cooperativity of formation above $T_F$ are shown in red
[blue].}
\label{fig:AB}
\end{figure}

As anticipated in section Theory, all $p_{i,j}$'s have monotonic
sigmoidal shapes which mainly reflect the sequence separation, $|i-j|$
and the native burial of each of the residues.  In general, each
contact is established at a different crossover temperature and with
different intensity \cite{hiv}. 
The data relative to the frequencies of native-contact formation is
conveniently summarised in the color-coded contact maps of
Fig.~\ref{fig:AB}.  A bright red color is used to highlight those
contacts with the largest crossover temperatures above $T_F$, see
Fig.~\ref{fig:AB}a, or highest intensity in Fig.~\ref{fig:AB}b.  Both
these intuitive notions can be used to identify the key folding
contacts.  The inspection of Fig.~\ref{fig:AB} reveals that several
kinetic bottlenecks (red regions) are located three-four contacts
downstream the three $\beta$-turns in HIV-1 PR.  In addition, the
formation of contacts around residues $84$ and $30$, despite being so
far away along the sequence, appears to be a crucial folding stage
since it allows the collapse of the individual secondary structure
motifs.  It is striking that these results make an excellent parallel
with those of Ref.~\cite{hiv}, where long and delicate MD simulations
of the unfolding/refolding of HIV-1 PR were carried out using a much
more sophisticated energy-scoring function.  This provides a cross
validation for the robustness of the results obtained both in the
stochastic and the present, analytic, scheme.  The emphasis is on the
exactness of the present approach that allows to determine easily the
$p_{i,j}$'s with an arbitrary accuracy.  The absence of stochastic
noise allows to compile Table~\ref{tab:tab1} which shows the top
contacts ranked according to crossover temperature and
intensity. Sites that are known to cause drug resistance through
mutations are highlighted in boldface. It is apparent that a high
fraction of the top key folding contacts do, indeed, contain key
mutating sites. To test the significance of such matches we compare
the number of marked mutating sites contained in each column of Table
II with the number of those contained in a randomly compiled table.
We expect a random list of $t$ elements extracted among $N$, $m$ of
which are marked, to contain an average of $t m/N$ marked elements
with a square deviation of $tm(N-m)(N-t)/(N^2 (N-1))$.  For the case
of HIV-1 PR the total number of contacts (excluding consecutive
residues) within a cutoff radius of 6.5 \AA\ is $N=180$ and the number
of those which include at least one known mutating site is $m=60$.  An
analysis of the contacts of Table \ref{tab:tab1} (selected according
to crossover temperature or cooperativity of formation) shows that the
number of matches observed among the top sites tipically exceeds that
expected from a random choice by a standard deviation (the precise
amount depend of how many top ranking contacts are considered.
An alternative and apparently more stringent approach is to identify
independent groups of highly correlated contacts, and then search for
the key residues in each group. To a first approximation, the
correlated sets of interacting pairs may be identified with the
clusters in the contact map. This leads to define six main groups, the
three $\beta$-sheets, the helix and the two sets of long-range
contacts, around contacts 14-60 and 23-84, respectively (see
Fig.~\ref{fig:AB}).  The four contacts in each group with the highest
intensity of formation above $T_F$ are summarised in
Table~\ref{tab:tab4}.  Out of the 24 contacts, 12 of them involve a
key site, which is two standard deviations away from the number of
matches expected on a random basis ($7.9 \pm 2.1$).  Again, this
testifies both the reliability of the general scheme followed here and
also its robustness in the different possible implementations.

Interestingly, the results of Table~\ref{tab:tab4} account better than
those of Table \ref{tab:tab1} for the heterogeneous location of the key
folding sites. The emerging conclusion is that a complete description
of the crucial contacts can be obtained only by monitoring all the key
stages of the folding process. In standard MC and MD simulations of
protein unfolding/refolding, it is the simulated dynamics that reveals
which, and how many, delicate stages exists.  In the present approach,
the folding process is characterised analytically, thus the complete
set of folding bottlenecks follows from the study of distinct groups
of interrelated contacts.

Finally, we remark that the determination of the key contacts does not
uniquely provide the key folding sites, since two sites are involved
in each pairwise contact. This ambiguity can, in several cases, be
resolved either by selecting those sites that take part in several
crucial contacts, or by examining their distribution on the
three-dimensional native structure for clues that may help breaking
the ambiguity.

\section{Conclusions}

We have used an analytical technique to study and characterize the
folding process of globular proteins. This deterministic method allows
the automated identification of contacts involved in folding
rate-limiting steps. As a result, the whole folding process is
particularly sensitive to mutations occuring at sites involved in such
crucial contacts.  We test our scheme and its usefulness in pinpointing
the crucial sites by applying it to HIV-1 protease. For this enzyme,
extensive clinical trials have allowed the identification of several
sites involved in drug-resistance mutations.  Such sites have a
meaningful overlap with the key folding sites predicted by our scheme
with a modest computational effort compared to more sophisticated
stochastic simulations techniques. This indicates that the available
inhibiting drugs are quite effective since they can be eluded only by
mutations of the (sensitive) key sites of the protease.

The proposed approach to identify the crucial residues is quite
general and ought to be useful to identify the kinetic bottlenecks of
other viral enzymes of pharmaceutical interest, thus aiding the
development of novel effective inhibitors.

We expect to focus our future efforts on improving the present
approach by taking into account the propensities of different amino
acids to form contacting pairs. This limitation can be overcome by
introducing physically viable (attractive) pairwise interactions
\cite{Seno:98:prl,Sippl95,Miyazawa96,Maiorov:92,stabloc}. In the
present approach this possibility was deliberately avoided to
highlight the influence of the native state topology alone on the
kinetic bottlenecks, irrespective of the different chemical nature and
strength of the effective amino acid interactions. We expect that the
inclusion of such effects, while not distorting the overall picture
presented here, may change the relative strength of spatially-close
contacts. This may improve the agreement between Table~\ref{tab:res}
and tables \ref{tab:tab1}-\ref{tab:tab4} by resolving those cases were
a site adjacent to a mutating one is selected.

We are indebted to Paolo Carloni for several illuminating discussions
and for having stimulated the present work.  This work was supported
by INFM, Murst Cofin2001.

\appendix
\section{Appendix}

In this appendix we discuss how the translational invariance of a
quadratic energy scoring function can be explicitly broken by fixing
the center of mass of the system in the origin. The constrained
partition function is written as:

\begin{equation}
{\cal Z}= \int \prod_{i=1}^N d^3 x_i\  e^{-1/2 \sum_{i,j} {\bf
x}_i\, A_{ij}\, {\bf x}_j}\, \delta^3 (\sum_i {\bf x}_i)
\label{eq:Zcons}
\end{equation}

\noindent where the matrix $A$ incorporates the quadratic dependence
of $H_0$ in eqn (\ref{eq:H_0}) from the space co-ordinates (and also
includes the $1/T$ factor to yield the usual Boltzmann weight). The
translational invariance of $H_0$ implies that $A$ satisfies the
property: $\sum_j A_{ij}=0$, which amounts to say that the uniform
vector, ${\bf v}_1 \equiv N^{-{1 \over 2}} \, (1,1,1,1...,1)$ is an
eigenvector of $A$ with eigenvalue $\lambda_1 =0$. We assume that
$H_0$ is invariant only for the simultaneous translation of all the
coordinates, $\{{\bf x}_i\}$. In this case all other eigenvalues,
$\{\lambda_{i>1}\}$ are strictly positive and the corresponding
eigenvectors ${\bf v}_{i>1}$ are all orthogonal to the zero mode ${\bf
v}_1$.

By rewriting the Dirac-delta constraint as
\begin{equation}
\delta^3({\bf z}) = \lim_{c \to \infty} \left({ c \over 2
\pi}\right)^{3 \over 2} e^{-c\ 
{\bf z} \cdot {\bf z} / 2}
\end{equation}

\noindent the partition function takes on the form ${\cal Z}= \lim_{c
\to \infty} {\cal Z}_c$, where
\begin{equation}
{\cal Z}_c = \left({ c \over 2 \pi}\right)^{3 \over 2}\ \int
\prod_{i=1}^N d^3 x_i\  e^{-1/2 \sum_{i,j} {\bf x}_i\, A^\prime _{ij} \, {\bf x}_j}\ ,
\end{equation}

\noindent where $A^\prime_{ij} = A_{ij} + c $. It is straightforward
to see that $A^\prime$ admits the same eigenvectors of $A$. Only the
zero mode eigenvalue will change from zero to $c\, N$, while all
others will be unmodified. Upon performing the Gaussian integrations
in ${\cal Z}_c$ we obtain:

\begin{eqnarray}
{\cal Z}_c  &=& \left({ c \over 2 \pi}\right)^{3 \over 2} \, \left({2 \pi \over c N}\right)^{3 \over 2}\ \ \prod_{i=2}^N \left({2 \pi \over
\lambda_i}\right)^{3 \over 2}\nonumber \\
&=& {1 \over N^{3/2}}\, \left({ 2 \pi }\right)^{3 (N-1) \over 2} \ \prod_{i=2}^N
{\lambda_i}^{- {3\over 2}}\ .
\end{eqnarray}

\noindent This shows that ${\cal Z}_c$ is effectively independent of
$c$ and, therefore, the partition function ${\cal Z}$ simplifies to

\begin{equation}
{\cal Z}= N^{- {3\over 2}} \ \left({2 \pi}\right)^{3 (N-1) \over 2} \, \left({\det
\null^\prime {A}}\right)^{-{3 \over 2}} \ ,
\end{equation}

\noindent where the prime denotes that the determinant is calculated
omitting the zero mode eigenvalue.

\newpage
\begin{table}
\begin{small}
\begin{center}
\begin{tabular}{|l|l|}
Drug & Point Mutations \\ \hline \hline
RTN \protect\cite{Molla,Marko} &20,33,35,36,46,54,63,71,82,84,90\\
NLF \protect\cite{patick} &30,{46},{63},71,{77},{84},\\
IND \protect\cite{condra,tisdale} &{10},{32},{46},{63},71,82,{84} \\
SQV \protect\cite{condra,tisdale,jacob}&{10},{46},48,{63},71,82,{84},90\\
APR \protect\cite{apr} &{46},{63},82,{84}\\
\end{tabular}
\end{center}
\end{small}
\caption{Mutations in the protease associated with FDA-approved drug
resistance \protect\cite{BIOCH88}.}
\label{tab:res}
\end{table}

\begin{table}
\begin{center}
\begin{tabular}{|l|l|}
Crossover Temperature & Cooperativity   \\  \hline \hline
  25  -  86  &              14 -  66       \\
  28  -  86  &              14  -  64       \\
  58  -  76  &             {\bf 10}  -  23  \\
  58  - {\bf 77}  &         14 -   65       \\
  57  - {\bf 77}  &         13  -  66       \\
  13  -  66  &              12  -  66       \\
 {\bf 30}  -  86  &         87  -  91       \\
 {\bf 32}  - {\bf 84}  &    13  -  65       \\
 {\bf 32}  -  76  &         23  - {\bf 84}  \\
  29  -  86  &             {\bf 10}  -   22 \\
  31  - {\bf 84}  &         56  - {\bf 77}  \\
  23  - {\bf 84}  &         57  - {\bf 77}  \\
  14  -  66  &              23  -  83       \\
  25  -  85  &              22  - {\bf 84}  \\
  14  -  65  &              57  -  78       \\
  45  -  56  &              86  -  89       \\
  89  -  91  &              34  -  78       \\
  13  -  65  &              58  - {\bf 77}  \\
  87  -  89  &             {\bf 30}  -  88  \\
 {\bf 84}  -  86  &        {\bf 32}  -   75 \\
  56  -  58  &             {\bf 32}  -   76 \\
  25  - {\bf 84}  &         31  -  76       \\
  86  -  88  &              42  -  58       \\
  64  - {\bf 71}  &        {\bf 90}   -  94 \\
  57  -  76  &              87  - {\bf 90}  \\
\end{tabular}
\end{center}
\caption{The top contacts ranked according to the crossover
 temperature (first column) and cooperativity of formation above $T_F$
(second column) }
\label{tab:tab1}
\end{table}

\begin{table}
\begin{center}
\begin{tabular}{|l|l|}
Bottlenekcs & Key Contacts \\ \hline \hline
$\beta_1$ &  {\bf 10} - 23 \\
$\beta_1$ &  {\bf 10} - 22 \\
$\beta_1$ &  14 - {\bf 20} \\
$\beta_1$ &  12 - {\bf 20} \\ \hline
$\beta_2$ &  42 - 58 \\
$\beta_2$ &  45 - 58 \\
$\beta_2$ &  43 - 58 \\
$\beta_2$ &  43 - 57 \\ \hline
$\beta_3$ &  56 - {\bf 77}\\
$\beta_3$ &  57 - {\bf 77}\\
$\beta_3$ &  58 - {\bf 77}\\
$\beta_3$ &  57 - 76\\ \hline
Other1 &  14 - 66 \\
Other1 &  14 - 64 \\
Other1 &  14 - 65 \\
Other1 &  13 - 66 \\  \hline
Other2 &  23 - {\bf 84} \\
Other2 &  23 - 83 \\
Other2 &  22 - {\bf 84} \\
Other2 &  {\bf 30} - 88 \\ \hline
Helix &   87 - 91  \\
Helix &   86 - 89  \\
Helix & {\bf 90} - 94  \\
Helix &  87 - {\bf 90} \\
\end{tabular}
\end{center} \caption{The four contacts with the highest cooperativity
of formation above $T_F$ for each of the six clusters of the contact
map.}
\label{tab:tab4}
\end{table}

\begin{thebibliography}{10}
\expandafter\ifx\csname bibnamefont\endcsname\relax
  \def\bibnamefont#1{#1}\fi
\expandafter\ifx\csname bibfnamefont\endcsname\relax
  \def\bibfnamefont#1{#1}\fi
\expandafter\ifx\csname url\endcsname\relax
  \def\url#1{\texttt{#1}}\fi
\expandafter\ifx\csname urlprefix\endcsname\relax\def\urlprefix{URL }\fi
\providecommand{\bibinfo}[2]{#2}
\providecommand{\eprint}[2][]{\url{#2}}

\bibitem{Fersht95}
\bibinfo{author}{\bibfnamefont{A.~R.} \bibnamefont{Fersht}},
  \bibinfo{journal}{Proc. Natl. Acad. Sci. USA} \textbf{\bibinfo{volume}{92}},
  \bibinfo{pages}{10869} (\bibinfo{year}{1995}).

\bibitem{sh3A}
\bibinfo{author}{\bibfnamefont{J.~C.} \bibnamefont{Martinez}} \bibnamefont{and}
  \bibinfo{author}{\bibfnamefont{L.}~\bibnamefont{Serrano}},
  \bibinfo{journal}{Nature Struct. Biol.} \textbf{\bibinfo{volume}{6}},
  \bibinfo{pages}{1010} (\bibinfo{year}{1999}).

\bibitem{sh3B}
\bibinfo{author}{\bibfnamefont{D.~S.} \bibnamefont{Riddle}},
  \bibinfo{author}{\bibfnamefont{V.~P.} \bibnamefont{Grantcharova}},
  \bibinfo{author}{\bibfnamefont{J.~V.} \bibnamefont{Santiago}},
  \bibinfo{author}{\bibfnamefont{E.}~\bibnamefont{Alm}},
  \bibinfo{author}{\bibfnamefont{I.}~\bibnamefont{Ruczinski}},
  \bibnamefont{and} \bibinfo{author}{\bibfnamefont{D.}~\bibnamefont{Baker}},
  \bibinfo{journal}{Nature Struct. Biol.} \textbf{\bibinfo{volume}{6}},
  \bibinfo{pages}{1016} (\bibinfo{year}{1998}).

\bibitem{Chiti99}
\bibinfo{author}{\bibfnamefont{F.}~\bibnamefont{Chiti}},
  \bibinfo{author}{\bibfnamefont{N.}~\bibnamefont{Taddei}},
  \bibinfo{author}{\bibfnamefont{P.~M.} \bibnamefont{White}},
  \bibinfo{author}{\bibfnamefont{M.}~\bibnamefont{Bucciantini}},
  \bibinfo{author}{\bibfnamefont{F.}~\bibnamefont{Magherini}},
  \bibinfo{author}{\bibfnamefont{M.}~\bibnamefont{Stefani}}, \bibnamefont{and}
  \bibinfo{author}{\bibfnamefont{C.~M.} \bibnamefont{Dobson}},
  \bibinfo{journal}{Nature Struct. Biol.} \textbf{\bibinfo{volume}{6}},
  \bibinfo{pages}{1005} (\bibinfo{year}{1999}).

\bibitem{Plaxco}
\bibinfo{author}{\bibfnamefont{K.~W.} \bibnamefont{Plaxco}},
  \bibinfo{author}{\bibfnamefont{K.~T.} \bibnamefont{Simons}},
  \bibnamefont{and} \bibinfo{author}{\bibfnamefont{D.}~\bibnamefont{Baker}},
  \bibinfo{journal}{J. Mol. Biol.} \textbf{\bibinfo{volume}{277}},
  \bibinfo{pages}{985} (\bibinfo{year}{1998}).

\bibitem{Miche99b}
\bibinfo{author}{\bibfnamefont{C.}~\bibnamefont{Micheletti}},
  \bibinfo{author}{\bibfnamefont{J.~R.} \bibnamefont{Banavar}},
  \bibinfo{author}{\bibfnamefont{A.}~\bibnamefont{Maritan}}, \bibnamefont{and}
  \bibinfo{author}{\bibfnamefont{F.}~\bibnamefont{Seno}},
  \bibinfo{journal}{Phys. Rev. Lett.} \textbf{\bibinfo{volume}{82}},
  \bibinfo{pages}{3372} (\bibinfo{year}{1999}).

\bibitem{M00}
\bibinfo{author}{\bibfnamefont{A.}~\bibnamefont{Maritan}},
  \bibinfo{author}{\bibfnamefont{C.}~\bibnamefont{Micheletti}},
  \bibnamefont{and} \bibinfo{author}{\bibfnamefont{J.~R.}
  \bibnamefont{Banavar}}, \bibinfo{journal}{Phys. Rev. Lett.}
  \textbf{\bibinfo{volume}{84}}, \bibinfo{pages}{3009} (\bibinfo{year}{2000}).

\bibitem{baker}
\bibinfo{author}{\bibfnamefont{E.}~\bibnamefont{Alm}} \bibnamefont{and}
  \bibinfo{author}{\bibfnamefont{D.}~\bibnamefont{Baker}},
  \bibinfo{journal}{Proc. Natl. Acad. Sci. USA} \textbf{\bibinfo{volume}{96}},
  \bibinfo{pages}{11305} (\bibinfo{year}{1999}).

\bibitem{Clem}
\bibinfo{author}{\bibfnamefont{C.}~\bibnamefont{Clementi}},
  \bibinfo{author}{\bibfnamefont{H.}~\bibnamefont{Nymeyer}}, \bibnamefont{and}
  \bibinfo{author}{\bibfnamefont{J.~N.} \bibnamefont{Onuchic}},
  \bibinfo{journal}{J. Mol. Biol.} \textbf{\bibinfo{volume}{298}},
  \bibinfo{pages}{937} (\bibinfo{year}{2000}).

\bibitem{cieplak}
\bibinfo{author}{\bibfnamefont{T.~X.} \bibnamefont{Hoang}} \bibnamefont{and}
  \bibinfo{author}{\bibfnamefont{M.}~\bibnamefont{Cieplak}},
  \bibinfo{journal}{J. Chem. Phys.} \textbf{\bibinfo{volume}{113}},
  \bibinfo{pages}{8319} (\bibinfo{year}{2000}).

\bibitem{bakernature}
\bibinfo{author}{\bibfnamefont{D.~A.} \bibnamefont{Baker}},
  \bibinfo{journal}{Nature} \textbf{\bibinfo{volume}{405}}, \bibinfo{pages}{39}
  (\bibinfo{year}{2000}).

\bibitem{Funnel2}
\bibinfo{author}{\bibfnamefont{P.~G.} \bibnamefont{Wolynes}},
  \bibinfo{author}{\bibfnamefont{J.~N.} \bibnamefont{Onuchic}},
  \bibnamefont{and}
  \bibinfo{author}{\bibfnamefont{D.}~\bibnamefont{Thirumalai}},
  \bibinfo{journal}{Science} \textbf{\bibinfo{volume}{267}},
  \bibinfo{pages}{1619} (\bibinfo{year}{1995}).

\bibitem{Jackson}
\bibinfo{author}{\bibfnamefont{S.~E.} \bibnamefont{Jackson}},
  \bibinfo{journal}{Folding and Design} \textbf{\bibinfo{volume}{3}},
  \bibinfo{pages}{R81} (\bibinfo{year}{1998}).

\bibitem{hiv}
\bibinfo{author}{\bibfnamefont{F.}~\bibnamefont{Cecconi}},
  \bibinfo{author}{\bibfnamefont{C.}~\bibnamefont{Micheletti}},
  \bibinfo{author}{\bibfnamefont{P.}~\bibnamefont{Carloni}}, \bibnamefont{and}
  \bibinfo{author}{\bibfnamefont{A.}~\bibnamefont{Maritan}},
  \bibinfo{journal}{Proteins: Structure Function and Genetics}
  \textbf{\bibinfo{volume}{43}}, \bibinfo{pages}{365} (\bibinfo{year}{2001}).

\bibitem{Settanni}
\bibinfo{author}{\bibfnamefont{G.}~\bibnamefont{Settanni}},
  \bibinfo{author}{\bibfnamefont{C.}~\bibnamefont{Cattaneo}}, \bibnamefont{and}
  \bibinfo{author}{\bibfnamefont{A.}~\bibnamefont{Maritan}},
  \bibinfo{journal}{Biophys. J.} \textbf{\bibinfo{volume}{80}},
  \bibinfo{pages}{2935} (\bibinfo{year}{2001}).

\bibitem{gaussian}
\bibinfo{author}{\bibfnamefont{C.}~\bibnamefont{Micheletti}},
  \bibinfo{author}{\bibfnamefont{J.}~\bibnamefont{Banavar}}, \bibnamefont{and}
  \bibinfo{author}{\bibfnamefont{A.}~\bibnamefont{Maritan}},
  \bibinfo{journal}{Phys. Rev. Lett.} \textbf{\bibinfo{volume}{87}},
  \bibinfo{pages}{DOI:088102} (\bibinfo{year}{2001}).

\bibitem{condra}
\bibinfo{author}{\bibfnamefont{J.~H.} \bibnamefont{{Condra et~al.}}},
  \bibinfo{journal}{Nature} \textbf{\bibinfo{volume}{374}},
  \bibinfo{pages}{569} (\bibinfo{year}{1995}).

\bibitem{flory}
\bibinfo{author}{\bibfnamefont{P.~J.} \bibnamefont{Flory}},
  \bibinfo{journal}{J. Am. Chem. Soc} \textbf{\bibinfo{volume}{78}},
  \bibinfo{pages}{5222} (\bibinfo{year}{1956}).

\bibitem{chan90}
\bibinfo{author}{\bibfnamefont{H.~S.} \bibnamefont{Chan}} \bibnamefont{and}
  \bibinfo{author}{\bibfnamefont{K.~A.} \bibnamefont{Dill}},
  \bibinfo{journal}{J. Chem. Phys.} \textbf{\bibinfo{volume}{92}},
  \bibinfo{pages}{3118} (\bibinfo{year}{1990}).

\bibitem{goddard}
\bibinfo{author}{\bibfnamefont{D.~A.} \bibnamefont{Debe}} \bibnamefont{and}
  \bibinfo{author}{\bibfnamefont{W.~A.} \bibnamefont{{Goddard III}}},
  \bibinfo{journal}{J. Mol. Biol.} \textbf{\bibinfo{volume}{294}},
  \bibinfo{pages}{619} (\bibinfo{year}{1999}).

\bibitem{CT2}
\bibinfo{author}{\bibfnamefont{C.~J.} \bibnamefont{Camacho}} \bibnamefont{and}
  \bibinfo{author}{\bibfnamefont{D.}~\bibnamefont{Thirumalai}},
  \bibinfo{journal}{Proc. Natl. Acad. Sci. USA} \textbf{\bibinfo{volume}{92}},
  \bibinfo{pages}{1277} (\bibinfo{year}{1995}).

\bibitem{Go}
\bibinfo{author}{\bibfnamefont{N.}~\bibnamefont{Go}} \bibnamefont{and}
  \bibinfo{author}{\bibfnamefont{H.~A.} \bibnamefont{Scheraga}},
  \bibinfo{journal}{Macromolecules} \textbf{\bibinfo{volume}{9}},
  \bibinfo{pages}{535} (\bibinfo{year}{1976}).

\bibitem{Klo99}
\bibinfo{author}{\bibfnamefont{A.}~\bibnamefont{Kloczkowski}} \bibnamefont{and}
  \bibinfo{author}{\bibfnamefont{R.~L.} \bibnamefont{Jernigan}},
  \bibinfo{journal}{Comp. Theor. Pol. Sci.} \textbf{\bibinfo{volume}{9}},
  \bibinfo{pages}{285} (\bibinfo{year}{1999}).

\bibitem{bah97}
\bibinfo{author}{\bibfnamefont{I.}~\bibnamefont{Bahar}},
  \bibinfo{author}{\bibfnamefont{A.~R.} \bibnamefont{Atilgan}},
  \bibnamefont{and} \bibinfo{author}{\bibfnamefont{B.}~\bibnamefont{Erman}},
  \bibinfo{journal}{Folding and Design} \textbf{\bibinfo{volume}{2}},
  \bibinfo{pages}{173} (\bibinfo{year}{1997}).

\bibitem{hiv99}
\bibinfo{author}{\bibfnamefont{I.}~\bibnamefont{Bahar}},
  \bibinfo{author}{\bibfnamefont{B.}~\bibnamefont{Erman}},
  \bibinfo{author}{\bibfnamefont{R.~L.} \bibnamefont{Jernigan}},
  \bibinfo{author}{\bibfnamefont{A.~R.} \bibnamefont{Atilgan}},
  \bibnamefont{and} \bibinfo{author}{\bibfnamefont{D.~G.}
  \bibnamefont{Covell}}, \bibinfo{journal}{Journal of Molecular Biology}
  \textbf{\bibinfo{volume}{285}}, \bibinfo{pages}{1023} (\bibinfo{year}{1999}).

\bibitem{kes00}
\bibinfo{author}{\bibfnamefont{O.}~\bibnamefont{Keskin}},
  \bibinfo{author}{\bibfnamefont{I.}~\bibnamefont{Bahar}}, \bibnamefont{and}
  \bibinfo{author}{\bibfnamefont{R.~L.} \bibnamefont{Jernigan}},
  \bibinfo{journal}{Biophysical Journal} \textbf{\bibinfo{volume}{78}},
  \bibinfo{pages}{2093} (\bibinfo{year}{2000}).

\bibitem{ani01}
\bibinfo{author}{\bibfnamefont{A.~R.} \bibnamefont{Atilgan}},
  \bibinfo{author}{\bibfnamefont{S.~R.} \bibnamefont{Durell}},
  \bibinfo{author}{\bibfnamefont{R.~L.} \bibnamefont{Jernigan}},
  \bibinfo{author}{\bibfnamefont{M.~C.} \bibnamefont{Demirel}},
  \bibinfo{author}{\bibfnamefont{O.}~\bibnamefont{Keskin}}, \bibnamefont{and}
  \bibinfo{author}{\bibfnamefont{I.}~\bibnamefont{Bahar}},
  \bibinfo{journal}{Biophysical Journal} \textbf{\bibinfo{volume}{80}},
  \bibinfo{pages}{505} (\bibinfo{year}{2001}).

\bibitem{BIOCH88}
\bibinfo{author}{\bibfnamefont{P.~J.} \bibnamefont{Ala}},
  \bibinfo{author}{\bibfnamefont{E.~E.} \bibnamefont{Huston}},
  \bibinfo{author}{\bibfnamefont{R.~M.} \bibnamefont{Klabe}},
  \bibinfo{author}{\bibfnamefont{P.~K.} \bibnamefont{Jadhav}},
  \bibinfo{author}{\bibfnamefont{P.~Y.~S.} \bibnamefont{Lam}},
  \bibnamefont{and} \bibinfo{author}{\bibfnamefont{C.~H.} \bibnamefont{Chang}},
  \bibinfo{journal}{Biochemistry} \textbf{\bibinfo{volume}{37}},
  \bibinfo{pages}{15042} (\bibinfo{year}{1998}).

\bibitem{Kaya1}
\bibinfo{author}{\bibfnamefont{H.}~\bibnamefont{Kaya}} \bibnamefont{and}
  \bibinfo{author}{\bibfnamefont{H.~S.} \bibnamefont{Chan}},
  \bibinfo{journal}{Phys. Rev. Lett.} \textbf{\bibinfo{volume}{85}},
  \bibinfo{pages}{4823} (\bibinfo{year}{2000}).

\bibitem{kaya2}
\bibinfo{author}{\bibfnamefont{H.}~\bibnamefont{Kaya}} \bibnamefont{and}
  \bibinfo{author}{\bibfnamefont{H.~S.} \bibnamefont{Chan}},
  \bibinfo{journal}{Proteins: Structure Function and Genetics}
  \textbf{\bibinfo{volume}{43}}, \bibinfo{pages}{523} (\bibinfo{year}{2001}).

\bibitem{CT}
\bibinfo{author}{\bibfnamefont{C.~J.} \bibnamefont{Camacho}} \bibnamefont{and}
  \bibinfo{author}{\bibfnamefont{D.}~\bibnamefont{Thirumalai}},
  \bibinfo{journal}{Proc. Natl. Acad. Sci. USA} \textbf{\bibinfo{volume}{90}},
  \bibinfo{pages}{6369} (\bibinfo{year}{1993}).

\bibitem{Funnel5}
\bibinfo{author}{\bibfnamefont{A.}~\bibnamefont{Sali}},
  \bibinfo{author}{\bibfnamefont{E.}~\bibnamefont{Shakhnovich}},
  \bibnamefont{and} \bibinfo{author}{\bibfnamefont{M.}~\bibnamefont{Karplus}},
  \bibinfo{journal}{Nature} \textbf{\bibinfo{volume}{369}},
  \bibinfo{pages}{248} (\bibinfo{year}{1994}).

\bibitem{Finkel}
\bibinfo{author}{\bibfnamefont{O.~V.} \bibnamefont{Galzitskaya}}
  \bibnamefont{and} \bibinfo{author}{\bibfnamefont{A.~V.}
  \bibnamefont{Finkelstein}}, \bibinfo{journal}{Proc. Natl. Acad. Sci. USA}
  \textbf{\bibinfo{volume}{96}}, \bibinfo{pages}{11299} (\bibinfo{year}{1999}).

\bibitem{karp}
\bibinfo{author}{\bibfnamefont{T.}~\bibnamefont{Lazaridis}} \bibnamefont{and}
  \bibinfo{author}{\bibfnamefont{M.}~\bibnamefont{Karplus}},
  \bibinfo{journal}{Science} \textbf{\bibinfo{volume}{278}},
  \bibinfo{pages}{1928} (\bibinfo{year}{1997}).

\bibitem{Seno:98:prl}
\bibinfo{author}{\bibfnamefont{F.}~\bibnamefont{Seno}},
  \bibinfo{author}{\bibfnamefont{C.}~\bibnamefont{Micheletti}},
  \bibinfo{author}{\bibfnamefont{A.}~\bibnamefont{Maritan}}, \bibnamefont{and}
  \bibinfo{author}{\bibfnamefont{J.~R.} \bibnamefont{Banavar}},
  \bibinfo{journal}{Phys. Rev. Lett.} \textbf{\bibinfo{volume}{81}},
  \bibinfo{pages}{2172} (\bibinfo{year}{1998}).

\bibitem{Sippl95}
\bibinfo{author}{\bibfnamefont{M.~J.} \bibnamefont{Sippl}},
  \bibinfo{journal}{Curr. Opin. Struct. Biol.} \textbf{\bibinfo{volume}{5}},
  \bibinfo{pages}{229} (\bibinfo{year}{1995}).

\bibitem{Miyazawa96}
\bibinfo{author}{\bibfnamefont{S.}~\bibnamefont{Miyazawa}} \bibnamefont{and}
  \bibinfo{author}{\bibfnamefont{R.~L.} \bibnamefont{Jernigan}},
  \bibinfo{journal}{J. Mol. Biol.} \textbf{\bibinfo{volume}{256}},
  \bibinfo{pages}{623} (\bibinfo{year}{1999}).

\bibitem{Maiorov:92}
\bibinfo{author}{\bibfnamefont{V.~N.} \bibnamefont{Maiorov}} \bibnamefont{and}
  \bibinfo{author}{\bibfnamefont{G.~M.} \bibnamefont{Crippen}},
  \bibinfo{journal}{J. Mol. Biol.} \textbf{\bibinfo{volume}{227}},
  \bibinfo{pages}{876} (\bibinfo{year}{1992}).

\bibitem{stabloc}
\bibinfo{author}{\bibfnamefont{C.}~\bibnamefont{Micheletti}},
  \bibinfo{author}{\bibfnamefont{F.}~\bibnamefont{Seno}},
  \bibinfo{author}{\bibfnamefont{J.~R.} \bibnamefont{Banavar}},
  \bibnamefont{and} \bibinfo{author}{\bibfnamefont{A.}~\bibnamefont{Maritan}},
  \bibinfo{journal}{Proteins: Structure Function and Genetics}
  \textbf{\bibinfo{volume}{42}}, \bibinfo{pages}{422} (\bibinfo{year}{2001}).

\bibitem{Molla}
\bibinfo{author}{\bibfnamefont{A.}~\bibnamefont{{Molla~et~al.}}},
  \bibinfo{journal}{Nat. Med.} \textbf{\bibinfo{volume}{2}},
  \bibinfo{pages}{760} (\bibinfo{year}{1996}).

\bibitem{Marko}
\bibinfo{author}{\bibfnamefont{M.}~\bibnamefont{{Markowitz~et~al.}}},
  \bibinfo{journal}{J. Virol.} \textbf{\bibinfo{volume}{69}},
  \bibinfo{pages}{701} (\bibinfo{year}{1995}).

\bibitem{patick}
\bibinfo{author}{\bibfnamefont{A.~K.} \bibnamefont{{Patick~et~al.}}},
  \bibinfo{journal}{Antimicrob. Agents Chemother.}
  \textbf{\bibinfo{volume}{40}}, \bibinfo{pages}{292} (\bibinfo{year}{1996}).

\bibitem{tisdale}
\bibinfo{author}{\bibfnamefont{M.}~\bibnamefont{{Tisdale~et~al.}}},
  \bibinfo{journal}{Antimicrob. Agents Chemother.}
  \textbf{\bibinfo{volume}{39}}, \bibinfo{pages}{1704} (\bibinfo{year}{1995}).

\bibitem{jacob}
\bibinfo{author}{\bibfnamefont{H.}~\bibnamefont{{Jacobsen~et~al.}}},
  \bibinfo{journal}{J. Infect. Dis.} \textbf{\bibinfo{volume}{173}},
  \bibinfo{pages}{1379} (\bibinfo{year}{1996}).

\bibitem{apr}
\bibinfo{author}{\bibfnamefont{P.}~\bibnamefont{Reddy}} \bibnamefont{and}
  \bibinfo{author}{\bibfnamefont{J.}~\bibnamefont{Ross}},
  \bibinfo{journal}{Formulary} \textbf{\bibinfo{volume}{34}},
  \bibinfo{pages}{567} (\bibinfo{year}{1999}).

\end{thebibliography}
\end{document}